\documentclass[12pt,preprint]{aastex}
\usepackage{emulateapj5}

\slugcomment{To appear in ApJ}

\newcommand\asca{{\it ASCA}}
\newcommand\sax{{\it BeppoSAX}}
\newcommand\chandra{{\it Chandra}}

\newcommand\xmm{{\it XMM-Newton}}

\newcommand\kev{{\rm~keV}}
\newcommand\ev{{\rm~eV}}
\newcommand\kms{\ifmmode {\rm~km\ s}^{-1} \else ~km s$^{-1}$\fi}
\newcommand\Hunit{\ifmmode {\rm~km\ s}^{-1}\ {\rm Mpc}^{-1}
        \else ~km s$^{-1}$ Mpc$^{-1}$\fi}
\newcommand\ctssec{\ifmmode {\rm~count\ s}^{-1} \else ~count s$^{-1}$\fi}
\newcommand\ergsec{\ifmmode {\rm~erg\ s}^{-1} \else
        ~erg s$^{-1}$\fi}
\newcommand\funit{\ifmmode {\rm~erg\ s}^{-1}\;{\rm cm}^{-2} \else
        ~ergs s$^{-1}$ cm$^{-2}$\fi}
\newcommand\phflux{\ifmmode {\rm~photon\ s}^{-1}\;{\rm cm}^{-2}
        \else   ~photon s$^{-1}$ cm$^{-2}$\fi}
\newcommand\efluxA{\ifmmode {\rm~erg\ s}^{-1}\;{\rm cm}^{-2}\;{\rm
        \AA}^{-1} \else ~erg s$^{-1}$ cm$^{-2}$ \AA$^{-1}$\fi}
\newcommand\efluxHz{\ifmmode {\rm~erg\ s}^{-1}\;{\rm cm}^{-2}\;{\rm
        Hz}^{-1} \else ~erg s$^{-1}$ cm$^{-2}$ Hz$^{-1}$\fi}
\newcommand\cc{\ifmmode {\rm~cm}^{-3} \else cm$^{-3}$\fi}
\newcommand\FWHM{\ifmmode {\rm~FWHM} \else ${\rm~FWHM}$\fi}
\newcommand\Msun{\ifmmode M_{\odot} \else $M_{\odot}$\fi}
\newcommand\Lsun{\ifmmode L_{\odot} \else $L_{\odot}$\fi}

\newcommand\hbeta{\ifmmode {\rm H}\beta \else H$\beta$\fi}
\newcommand\Kalpha{\ifmmode {\rm K}\alpha \else K$\alpha$\fi}
\newcommand\NH{\ifmmode N_{\rm H} \else N$_{\rm H}$\fi}
\usepackage{graphicx}
\usepackage{here}

\begin{document}

\title{Iron K$\alpha$ emission from the low-luminosity Active Galaxies M81 and  NGC 4579}

\author{G. C. Dewangan\altaffilmark{1,2}, R. E. Griffiths\altaffilmark{1}, T. Di Matteo\altaffilmark{3}, \& N. J. Schurch\altaffilmark{1}} 
\altaffiltext{1}{Department of Physics, Carnegie Mellon University, 5000 Forbes Avenue, Pittsburgh, PA 15213 USA}
\altaffiltext{2}{email: gulabd@cmu.edu}
\altaffiltext{3}{Max-Planck-Institut f{\" u}r Astrophysik, Karl-Schwarzschild-St
r.~1, 85740 Garching bei M{\" u}nchen, Germany}

\begin{abstract}
  We report on \xmm{} spectroscopy of the
  low-luminosity active galaxies (LLAGN) M81 and NGC~4579 both of
  which have known black hole masses and well-sampled spectral energy
  distributions (SED). The iron K$\alpha$ line profiles from both the
  LLAGN can be described in terms of two components -- a narrow line
  at $6.4\kev$ and a moderately broad line (FWHM $\sim 2 \times
  10^{4}\kms$) arising from highly ionized, He-like or H-like species
  ($E \sim 6.8\kev$). We interpret the broad lines arising from an
  accretion disk the inner edge of which is restricted to large radii
  ($r_{in} \sim 100 r_g$). However, the Eddington ratio,
  ${L}/{L_{Edd}}$, of these sources, is 3--4 orders of magnitude lower
  than that required to photo-ionize a cold disk to He-like iron. 
  We suggest that the
  lines can be explained as collisionally ionized X-ray lines arising
  from the transition region between a hot (radiatively inefficient)
  flow in the inner regions and a cold disk outside $r \sim 100r_g$.
  The accretion flow geometry probed by our {\it XMM-Newton}
  observations is consistent with the truncated disk models proposed
  to explain the SED of LLAGNs. 

\end{abstract}


\section{Introduction}

Active galactic nuclei (AGN) typically display a `big blue bump' 
in their optical/UV spectra
interpreted as blackbody emission arising from optically thick,
geometrically thin accretion disks. Their hard X-ray power-law
spectrum is thought to be produced by Comptonization of the soft disk
photons in a hot corona above the accretion disk (e.g.; Haardt \&
Maraschi 1991). Low luminosity AGN (LLAGN), emitting well below 
their Eddington limit (${L}/{{L}_{Edd}} \le 0.01$), typically 
lack the `big blue bump' (e.g., Ho 1999; Di Matteo et al.~2003 for 
the case of M87) and for this
reason it has been proposed that in these objects the accretion disk
consists of of two zones: an outer thin disk that extends from some
large radius down to a transition radius and an inner low-radiative
efficiency accretion flow close to the black hole.

Fluorescence Fe K$\alpha$ line emission can be a very powerful probe for
the accretion flow geometry around black holes. Such lines have been
interpreted as being produced through X-ray irradiation of the accretion
disk; therefore the physical width of the line can serve as a trace of
the inner accretion disk extent.  Observations of relativistically
broadened iron K$\alpha$ lines from some Seyfert galaxies
(e.g., MCG-6-30-15 -- Wilms et al 2001; Fabian et al. 2002), which
require that the disk must be irradiated from an X-ray source
within $6r_g$, 
suggests that at least in this case the
accretion disk extends to the last stable orbit. It is interesting to
note that MCG-6-30-15 (FWHM(H$\alpha$)$\sim 2200\kms$; Sulentic et al.
1998) is like narrow-line Seyfert~1 galaxies (NLS1), which are thought to 
 be accreting at relatively high accretion rates.  

In contrast, Lasota et al. (1996) suggested that the accretion disk in
most LLAGN must be truncated. Gammie, Narayan, and Blandford (1999)
have shown the broadband spectrum of NGC~4258 can be explained by
emission from a geometrically thin accretion disk + ADAF model with a
transition radius at $\sim 10-100$ Schwarzschild radii ($r_S$). On the basis of their
optical/UV continuum spectra, Quataert et al. (1999), have provided
further evidence for such accretion flow geometry and showed that the
spectral energy distributions of M~81 and NGC~4579 can be explained by
an ADAF + disk with a transition radius at $\simeq 100 r_S$.

One prediction from any such models, as discussed in
Quataert et al (1999), is the lack of a relativistically broadened
iron K$\alpha$ line in the X-ray spectrum of these objects. The 
presence, or absence, of this feature in X-ray spectra from 
\chandra{} and \xmm{} can therefore be used to test the accretion 
flow geometry proposed for LLAGN. Specifically, any detection of 
relativistically broad iron K$\alpha$ line would strongly argue 
against truncated disk models.

In this paper, we utilize \xmm{} observations of 
two of the best studied LLAGN, NGC~4579 and M~81, to investigate whether the
observed iron K$\alpha$ profiles are consistent with 
truncated disk models. Both NGC~4579 and M~81 show broad H$\alpha$ line and 
are classified as type 1 AGN (Barth et al. 2001; Filippenko \& Sargent 1985). 

\section{Observation and Data Reduction}
\xmm{} observed NGC~4579 and M~81 on 12 June 2003 and 22-23 April
2001 respectively, using he European Photon Imaging Camera (EPIC),
the reflection grating spectrometer (RGS), and the optical monitor
(OM) instruments.  The EPIC PN \citep{Struderetal01} and MOS
\citep{Turneretal01} cameras were operated in full frame mode using
the thin filter for the observation of NGC~4579. 
 For the observation of M 81, the EPIC PN 
 was operated in small window, the MOS1 camera was operated in fast
 uncompressed mode and the MOS2 camera was operated in full frame
 mode, with the medium filter in place for all three detectors.
We present the spectral data obtained from the  EPIC cameras for
NGC 4579 and the spectral data from the EPIC PN and MOS2 camera for M~81.
The EPIC PN exposures were 22ks and 130ks for NGC~4579 and M~81, respectively.

The raw events were processed and filtered using the most recent
updated calibration database and analysis software ({\tt SAS v5.4.1})
available in August 2003. Events in the bad pixels file and those
adjacent pixels were discarded.  Only events with pattern $0-4$ (single and
double) for the PN and $0-12$ for the MOS were selected.  
Light curves extracted from source free
regions were examined for flaring particle background and affected
exposure intervals were excluded from the rest of the analysis.  This
resulted in `good' exposure time of $15.2{\rm~ks}$ and $ {\rm~ks}$ for
NGC~4579 and $71.9{\rm~ks}$ M~81, respectively.

Source spectra were extracted from a circular region of radius
$50\arcsec$ for both NGC~4579 and M~81.  Associated background spectra
were extracted from source free regions. The total number of source
photons, detected with the PN is $4.1\times 10^4$ for NGC 4579 and
$5.44\times10^{5}$ for M~81.  The source spectra were grouped to a
minimum of 20 counts per spectral channel with appropriate sampling of
the data ($\sim 3$ channels per spectral resolution elements) and were
analyzed using {\tt XSPEC 11.2.0} (Arnaud 1996). All the errors quoted 
below were calculated at $90\%$ confidence level for one interesting parameter.

\section{Detection of complex iron line profiles}
In the following analysis, only the EPIC data above $2.5\kev$ are
examined, the results presented in this paper are not critical to the
soft X-ray spectrum.  The soft X-ray spectrum of M~81 obtained from
the RGS observation has been studied by Page et al. (2003b).  For each
object, at first the MOS and PN spectra were fitted separately to
check for the uncertainties due to cross-calibration problems. We
found generally good agreement between MOS2 and PN cameras ($\Delta
\Gamma \sim 0.03$) in the $2.5-10\kev$ bands. However, the MOS1 data results in
a flatter ($\Delta\Gamma \sim 0.15$) spectrum. These discrepancies are
consistent with that found by Molendi \& Sembay (2003).  Therefore we
present the spectral results obtained by fitting the same model
jointly to the PN and MOS2 data in the $2.5-12\kev$ band, and allowing 
the relative normalizations to vary. 

\subsection{NGC 4579}
We fit the PN and MOS2 spectrum of NGC 4579 jointly with a power-law
model modified by photoelectric absorption. This fit resulted in a
$\chi^2$ of $296.6$ for 253 degrees of freedom (dof). The best-fit
absorption column density is consistent with the Galactic column
density ($N_{H}=3.1\times10^{20}{\rm~cm^{-2}}$ along the line of sight
of NGC~4579. From this simple fit, at least two iron line components
are evident in the X-ray spectrum of NGC~4579.  Figure~\ref{f1} shows
the iron K$\alpha$ line profile of NGC~4579 as the ratio of the
observed PN data to the the best-fit continuum model (see below).
Addition of a narrow Gaussian line at $\sim 6.4\kev$ improved the fit
significantly ($\chi^2=267.1$ for 251 dof). However, excess emission
is seen at $\sim 6.7\kev$. The addition of a broad Gaussian line 
(model 1: power-law + narrow Gaussian + broad Gaussian)
improves the joint fit significantly ($\chi^2=242.5$ for $249$ dof),
over the best-fit single Gaussian line model and also results in a
better fit than the best-fit from a model consisting of an absorbed
power-law and two narrow Gaussian lines ($\chi^2=257.0$ for $249$
dof).  The best-fit parameters are $\Gamma=1.77_{-0.07}^{+0.06}$,
narrow Gaussian: $E_{K\alpha,n}=6.40\kev$ (fixed), $EW=106\ev$, broad
Gaussian: $E_{K\alpha,b}=6.79_{-0.13}^{+0.11}\kev$,
$\sigma=231_{-97}^{+102}\ev$, $EW=287\ev$ (see model 1 in Table~\ref{tab1}).

The observed flux of NGC~4579 in the $2.5-12\kev$ band is $3.9\times
10^{-12}{\rm~erg~cm^{-2}~s^{-1}}$, and the corresponding luminosity is
$1.2\times10^{40}{\rm~erg~s^{-1}}$ (for a distance of
$d=16.7{\rm~Mpc}$ as measured by Pierce \& Tully, 1998). An {\it
  ASCA} observation of NGC~4579 in 1995 showed a strong (EW $\sim
500\ev$), marginally broad, symmetrical, iron K$\alpha$ line profile
at $6.73\pm0.13\kev$, while the previous 1998 observation showed a
narrow line at $6.39\pm0.09\kev$ (EW $\sim 260\ev$) with no clear line
at $\sim 6.7\kev$. It appears that both the above components observed
with \asca{} are clearly detected in the current \xmm{} observation.

\subsection{M 81}
The PN and MOS2 spectra of M81 were also fit by a power-law
modified by photoelectric absorption. The best-fit absorption column
density is consistent with the Galactic value
($\NH=4.2\times10^{20}{\rm~cm^{-2}}$).  This simple fit is
statistically unacceptable ($\chi2$ of $397.5$ for 303 dof) due to
significant spectral features in the iron K$\alpha$ region (see
Fig.~\ref{f1}).  The complex line profile of M~81 is not well
described by a narrow Gaussian line with the power-law plus narrow Gaussian model 
resulting in a poor fit ($\chi^2=363.0$ for 303
dof). The addition of a second narrow Gaussian line at $\sim 6.7\kev$
improves the fit significantly ($\chi^2 = 318.3$ for 301 dof). Varying
the width of this line further improves the fit ($\chi^2 = 299.3$ for
300 dof). The best-fit parameters are $\Gamma_X=1.87\pm0.02$, narrow
Gaussian: $E_{K\alpha,n}=6.40\kev$ (fixed), $EW=31\ev$, broad Gaussian:
$E_{K\alpha,b}=6.79_{-0.07}^{+0.06}\kev$, $\sigma=188_{-57}^{+76}\ev$,
$EW=101\ev$.

The observed flux of M~81 in the $2.5-12\kev$ band is $1.2\times
10^{-11}{\rm~erg~cm^{-2}~s^{-1}}$, and the luminosity in the same band
is $1.8\times10^{40}{\rm~erg~s^{-1}}$ (for a distance of
$d=3.6{\rm~Mpc}$; Freedman et al. 1994).

The \xmm{} observations of M~81 confirm the previous \asca{} and
\sax{} results that detected an intrinsically broad or complex iron line
centered at $\sim6.7\kev$, with an equivalent width of
$\sim170\pm60\ev$ (Ishisaki et al.1996). Neither \asca{} nor \xmm{}
observations detected an absorption edge at $\sim 8.6\kev$, similar to
that observed previously with \sax{} (Pellegrini et al.~2000).

\section{Inferences from the iron lines}

Both NGC~4579 and M~81 show iron line emission with broad ($FWHM \sim
20000\kms$), symmetrical line profiles, in addition to narrow iron
K$\alpha$ lines at $6.4\kev$.

The broad iron K$\alpha$ lines cannot be easily interpreted as thermal
emission from the host galaxy because the inferred thermal broadening
from such a system is too small ($FWHM\sim 350\kms$ for a temperature
of $10^8{\rm~K}$) to explain the widths of the observed lines.  A
thermal origin for the lines is further ruled out by the variability
of the lines as observed by \asca{} (Terashima et al. 2000). 

The observed broad
line from NGC~4579 and M~81 may thus arise from the accretion disk, as
generally assumed in Seyfert galaxies. 
The broad Gaussian lines in the current models were replaced with the
disk-line model of Fabian et al. (1989) in order to test this
hypothesis. This model assumes a Schwarzschild geometry and a disk
emissivity that is a power-law as a function of disk radius ($\epsilon
\propto r^{-q}$).  The emissivity index was fixed at $q=2.5$, the
average value for Seyfert~1 galaxies derived by Nandra et al. (1997).
The outer disk radius was also fixed at $1000r_g$ and the disk
inclination angle $i$ was fixed at $30\deg$.  The free parameters of
the disk-line model were the line energy, the inner radius of the
accretion disk, and the line flux.  This model 
(model 2: power-law + narrow Gaussian + diskline) results in an
acceptable fit for both AGN; $\chi^2/dof=240.6/249$ for NGC~4579 and
$\chi^2/dof=296.3/300$ for M~81. The continuum and narrow Gaussian
line parameters did not change significantly (see Table~\ref{tab1}).

\begin{figure*}
  \centering
  \includegraphics[width=6cm,angle=-90]{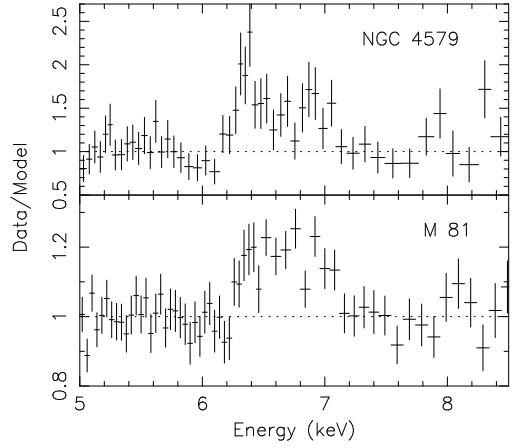}
\caption{Iron K$\alpha$ line profiles of NGC~4579 and M~81 shown as the ratio of the observed PN data and the best-fit continuum model (see text).
\label{f1}}
\end{figure*}

\begin{figure*}
\centering
\includegraphics[width=6cm,angle=-90]{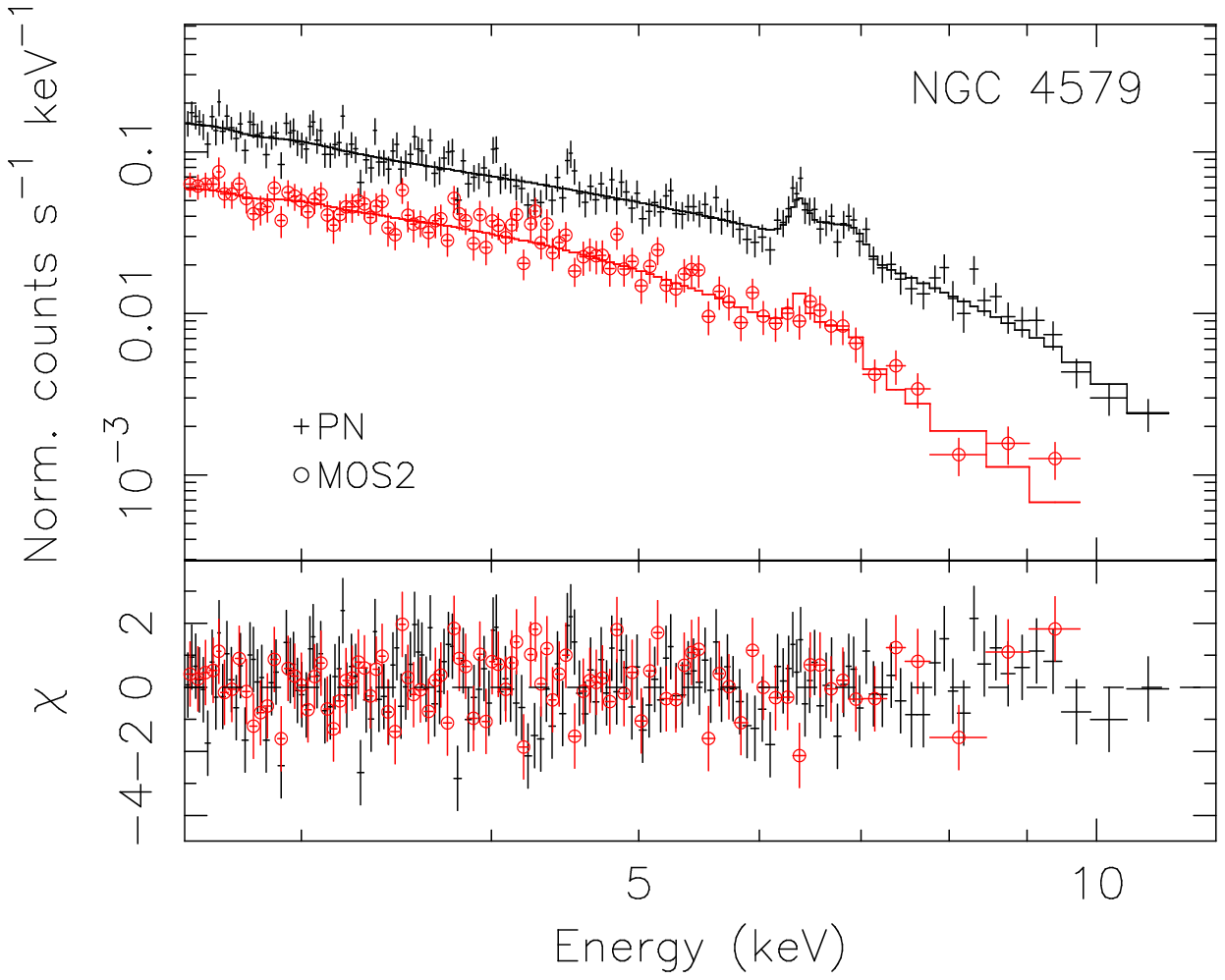}
\includegraphics[width=6cm,angle=-90]{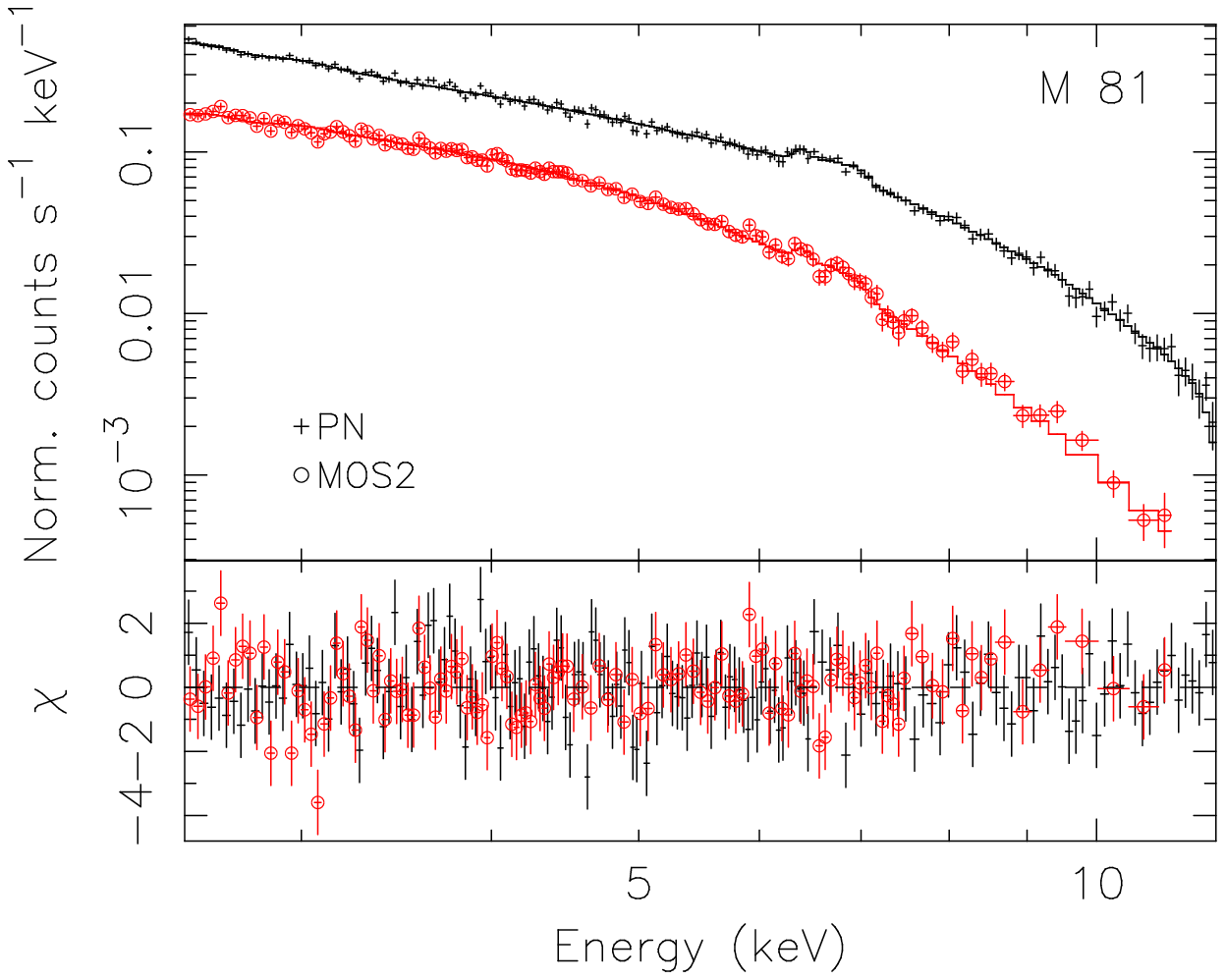}
\caption{The observed EPIC PN and MOS2 spectra of NGC~4579 and M~81, the best-fit models consisting of an absorbed power-law, a narrow Gaussian, and a diskline (see text). Lower panels show the deviations of the observed data from the best-fit models. 
\label{f2}}
\end{figure*}

Figure~\ref{f2} shows the observed PN data, the best-fit model (which
includes an absorbed power law a narrow Gaussian and a diskline) and the
deviations (in units of $\sigma$) for NGC~4579 and M~81. The iron
K$\alpha$ profiles, shown in Fig.~\ref{f1} as the ratio of observed
data and the best-fit continuum model, were obtained by setting the
normalizations of the narrow and broad K$\alpha$ lines to be zero. 
Only the ratios to the PN data are plotted in Fig.~\ref{f1} for clarity.
In order to evaluate the
dependence of the line properties on disk inclination, we also fit the
data for $i=60\deg$. This results in $E_{diskline} =
6.70_{-0.09}^{+0.09}\kev$; $r_{in} = 232_{-117}^{+490} r_g$ for
NGC~4579 and $E_{diskline} = 6.82_{-0.10}^{+0.05}\kev$;
$r_{in}>160r_g$ for M~81.
Fitting accretion disk models to the data suggest that the line
emitting region of the disk is truncated at an inner radius 
of about or greater than $100 r_g$ (the  narrow core
to the $6.4$ keV line suggests that this component
must also originate at a substantial distance from the black hole).

Our best-fit model does not show any
residuals indicating the presence of a substantial reflection continuum (from
either neutral of partially ionized matter from a disk close to the
black hole) in the spectral data of either M81 or NGC~4579 (see Fig.
2). The apparently lack of reflection continuum was also confirmed by
the detailed spectral analysis of the \sax{} observation of M81
(Pellegrini 2000).
It is therefore unlikely the the $6.7$ keV emission is  produced
by reflection from an ionized disk (see also \S 5.2).
To assess the possible origin of the $\sim 6.7\kev$ broad feature in 
the spectrum of NGC~4579  as the Compton reflection emission, 
we fitted a Compton reflection model ({\tt pexrav}; Magdziarz \&
Zdziarski 1995) along with the absorbed power-law and narrow Gaussian
line model to the PN spectrum of NGC~4579. The best-fit of this
reflection model results in an unrealistically  large relative reflection parameter,
$R=8.2_{-1.5}^{+8.9}$ (where $R=1$ corresponds to a solid angle of
$2\pi$ subtended by the reflector) and the fit worsens 
($\Delta \chi^2 = +20.7$ for an additional dof) in comparison to the best-fit 
obtained with the absorbed power law and double Gaussian line model (see Table~\ref{tab1}).
Thus the unphysical value of the reflection parameter and the
poor quality of the fit rule out 
 Compton reflection emission in place of the fluorescence iron
K$\alpha$ emission.

Finally, a combination of multiple narrow Gaussian lines was also tested as a
model to fit the complex iron line profiles observed in these sources.
Three narrow Gaussian emission lines were added to the absorbed power-law 
model with fixed line energies of
$6.4$, $6.7$ and $7.0\kev$, corresponding to cold, He-line, and H-like
iron.  The fits based on this model are statistically acceptable fits
(NGC~4579, $\chi^2=248.5$ for 250 dof; M~81, $\chi^2=304.1$ for 301
dof). These are worse fits compared to the diskline
models ($\Delta\chi^2=+7.9$ and $+7.8$ for NGC~4579 and M~81,
respectively) at a confidence level of $99.5\%$ based on an F-test. 
With the current generation of X-ray spectroscopy
satellites it is not possible to clearly distinguish between the multiple
($\ge 3$) narrow line and the disk-line models however, observations 
with, {\it Astro~E2}, and {\it Constellation-X}, with their correspondingly 
higher spectral resolution and sensitivity will be particularly important to 
further investigate the complexity of these iron line profiles.

\section{Origin of iron line complex}
Both NGC~4579 and M~81 show complex Iron K$\alpha$ features which are
well described by a combination of a narrow Gaussian and either a
broad Gaussian emission line, a disk-line, or a combination of narrow
emission lines from multiple ionization states of iron.

\subsection{The neutral, narrow iron K$\alpha$ line}
Despite the good spectral resolution of the EPIC cameras, the
line energies and the widths of the observed narrow K$\alpha$ lines
are not well constrained. The line parameters for NGC~4579 are
similar to the narrow lines commonly observed from Seyfert galaxies
with \chandra{} and \xmm{} (e.g., Yaqoob et al. 2001; Pounds et al.
2001), however, the strength of the narrow K$\alpha$ line is weaker in
M~81, in comparison to that of 'typical' Seyfert~1 galaxies (see e.g.
Page et al. 2003a). The observed narrow-lines from NGC~4579 and M~81
can be produced anywhere in the broad-line region, dusty torus, outer
regions of an standard accretion disk or a disk with a flat emissivity
law (Yaqoob et al. 2001,2003).  The weaker narrow line of M~81 may
suggest a smaller covering fraction of the material responsible for
the narrow line emission in NGC~4579 than that in M~81.

\subsection{The ionized, broad iron K$\alpha$ line} 

The widths and line center energies of the broad K$\alpha$ lines from
NGC~4579 and M~81 are remarkably similar ($E\sim 6.7-6.8\kev$; FWHM
$\sim 20000\kms$) suggestive of both an origin in similar regions the
AGN and a similarity in the ionization stages (He- or H-like iron) of
the emitting material.  The observed line energies are higher than
those generally observed from Seyfert 1. In general, only narrow-line
Seyfert~1 galaxies (NLS1s), with steep X-ray spectra, regularly show
strong, highly ionized iron K$\alpha$ lines (Dewangan 2002). In such
cases the lines are interpreted as being produced by reflection from
material in the disk photo-ionized by the strong X-ray continuum.  The
physical problem with explaining the lines in M81 and NGC~4579 in a
similar way is that we do not expect photoionization to be important
in these low luminosity sources.  For photo-ionized material, the
ionization parameter can be written as $\xi=\frac{L}{nR^2}$, where $L$
is the luminosity of ionizing photons, $n$ is the density of the
material, and $R$ is the distance of the material from the ionizing
source (Kallman \& McCray 1982). For standard thin disks, Matt et al.
(1993) have shown that the ionization parameter is strongly dependent
on the accretion rate relative to the Eddington rate and hence can be
rewritten as, $\xi \propto ({{L}/{L_{Edd}}})^3$.  In order to ionize
He-like iron, $\xi$ should be at least $\sim500$ or
${L}/{L_{Edd}}\gtrsim 0.2$ (Matt et al.  1993). This result is
consistent with the observation of highly ionized iron K$\alpha$ lines
from some NLS1s as they are thought to emit a higher fraction of their
Eddington luminosity (Pounds, Done, \& Osborne 1995). For NGC~4579,
the total nuclear luminosity of $\sim 5\times10^{41}\ergsec$ (Barth et
al.  1996) and black hole mass in the range of
$1.8-9.3\times10^7{\rm~M\odot}$ (Barth et al. 2001) correspond to
${L}/{L_{Edd}}\sim (0.5-2.2)\times10^{-4}$ while Laor (2003) reported
${L}/{L_{Edd}}\sim 1.35\times10^{-4}$. Ho et al.  (1996) found
${L}/{L_{Edd}}\sim (2-10)\times10^{-4}$ for M~81 which has a black
hole mass of $4 \times 10^{6} \Msun$.  Thus the ionizing luminosity of
both the LLAGNs NGC~4579 and M~81 are 3-4 orders of magnitude smaller
than that required for a photo-ionized accretion disk with He-like
iron. Clearly, the dominant ionization process in the disk of NGC~4579
and M~81 cannot be photo-ionization.

The widths of the ionized iron K$\alpha$ lines from NGC~4579 and M~81
(FWHM$\sim20000\kms$) are not as broad as the line widths predicted by
models of an accretion disk extending down to the last stable orbit.
Assuming Keplerian motion of the emitting material (i.e., $r\sim
({c^2}/{v_{FWHM}^2}) r_g$) the size of the broad iron K$\alpha$ line
emitting material is estimated to be $\sim150 r_g$ for NGC~4579 and
$\sim230 r_g$ for M~81, independent of black hole mass. Thus the
majority of iron K$\alpha$ photons are not produced near the last
stable orbit for a Schwarzschild ($r_{in}=6r_g$) or a Kerr black hole
($1.235r_g\le r_{in} \ 6r_g$). The same conclusion can be drawn from
the disk-line fits which resulted in $r_{in}=50_{-21}^{+111}r_g$ for
NGC~4579 and $65_{-27}^{+105}r_g$ for M~81.

Two possible scenarios that can explain the observed
moderate line widths and their $\sim6.7-6.8\kev$ centroid energy  are: 
($i$) the
accretion disk is fully ionized for $r<r_{in}$ so there are no lines
arising from within this region, or ($ii$) the accretion disk is truncated at
$r\sim r_{in}$. As discussed above, however, the accretion disk
cannot be photo-ionized  
by irradiation, and thus 
 must be intrinsically
hot. Both ($i$) and ($ii$) then imply that the inner regions of the
accretion flows in M~81 and NGC~4579 LLAGNs are likely to be
different from those of typical Seyfert galaxies. 

\subsection{The truncated disk model}
Our results are consistent with the suggestion that LLAGN have
accretion disks that are truncated at some large radii and have a hot
flow in the inner regions (Lasota et al. 1996; Narayan 1996; Gammie et
al.; Li \& Wang 2000). In particular, the inner radius $r_{in}\sim100
r_s$ of the accretion disk implied by the observed iron line width is
fully consistent with the position of the truncation radius deduced
from the modeling of the optical/UV spectra of M~81 and NGC~4579 by
Quataert et al. (1999; assuming a geometrically-thin accretion disk
and an optically thin, two-temperature ADAF in the inner regions).
Note also that in
this model, the continuum X-ray emission is produced in the inner
region of the hot flow (from $r=1 - 100 r_s$) and is due to
Comptonization of synchrotron emission from the flow.

\begin{figure*}
\centering
\includegraphics[width=6cm,angle=-90]{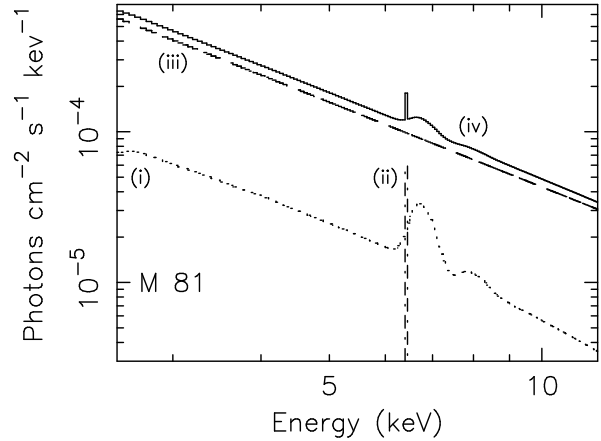}
\caption{The best-fit model consisting of (i) mekal plasma model, (ii) a narrow Gaussian, and (iii) a power-law derived from the joint fit to the PN and MOS2 spectra of M~81. The model is shown for the PN data only for clarity, the model for the MOS2 data differs in relative normalization by a small fraction ($\sim 0.02\%$).  The mekal model was convolved with a Gaussian of $\sigma\sim0.2\kev$ (see text)}.
\label{f3}
\end{figure*}

We suggest that the relatively broad $6.7$ keV lines may arise from
regions of the flow close to or at the transition radius, where the hot,
fully ionized flow starts cooling to connect to the outer optically
thick, geometrically thin region. Using the dynamical solutions of
the transition regions from a thin disk to an ADAF recently derived by
Manmoto et al. (2001) we can try to estimate the densities and
temperatures typical for this transition region.
In order for the hot flow to connect to a
standard thin disk the temperature and velocity of the gas must change
by orders of magnitude over a small transition region between the ADAF
and the thin disk. For a transition radius of $r_{tr} \sim 100 r_s$,
the temperature ($T \propto c_s^2$) scales as $\sim r^{-1}$ in the inner 
regions, and as
it approaches the transition radius, it then drops sharply from values
of $T\sim10^9$ K at $r\sim50 r_s$ to $\sim10^8$ K at $r\sim100 r_s$
and cooler beyond. As pressure balance has to be maintained, the
density of the flow increases sharply (by $\sim$2 orders of magnitude)
in the same range of radii, reaching values of $n_e\sim2\times10^{12}$
cm$^{-3}$ at $r\sim100 r_s$.  Even using the values of the flow
density derived above, photo-ionization in the transition region is
not likely to be important because of the low level of the ionizing
hard X-ray flux, in fact, at the transition region we find that the
ionization parameter $\xi\sim$1. Instead,
the observed He-like and H-like iron lines are likely to represent
collisionally ionized X-ray lines produced in a hot, $T\sim10^8$ K,
transition region. Specific calculations of the emission line profiles from
this region are not available at the moment and would be necessary
to fully explore this hypothesis. To
test this inference, therefore, we consider a simple 
thermal plasma model ({\tt mekal};
Liedahl, Osterheld, \& Goldstein (1995) and references therein) and
fit it in combination with a power-law continuum and a narrow Gaussian
emission line at $6.4\kev$ (fixed) to the PN and MOD spectra jointly.
The density of the plasma was fixed to $n=2\times10^{12}{\rm~cm^{-3}}$
(as derived above) and no intrinsic absorption is required. 
The mekal model was
convolved with a Gaussian in order to mimic the line broadening due to
Keplerian motion of the plasma at the transition region. The Gaussian
$\sigma$ was fixed at $0.2\kev$, similar to that derived for the broad
Gaussian component in the double Gaussian model for the iron line (see
Table~\ref{tab1}). The model provided a good fit to the data
($\chi^2/dof= 246.4/249$ for NGC~4579 and $298.0/296$ for M~81),
albeit somewhat worse than the models incorporating either a broad
Gaussian or diskline.  As an example, this model is shown in
Figure~\ref{f3} for the PN data of M~81. The best-fit parameters are
$\Gamma=1.73_{-0.11}^{+0.06}$, $kT=8.3_{-5.6}^{+7.1}\kev$ for NGC~4579
and $\Gamma=1.87\pm0.04$, $kT=8.5_{-3.0}^{+3.2}\kev$ for M~81.  The
best-fit plasma temperatures for both the AGN are similar to that
deduced above for the transition region. The power-law and the mekal
fluxes are $2.6\times10^{-12}$ and
$1.3\times10^{-12}{\rm~erg~cm^{-2}~s^{-1}}$ for NGC~4579 and
$1.0\times 10^{-11}{\rm~erg~cm^{-2}~s^{-1}}$ and $1.3\times
10^{-12}{\rm~erg~cm^{-2}~s^{-1}}$, respectively.

The success of this model in reproducing the observed iron K$\alpha$
profiles suggests that the observed lines could be produced
by collisional excitation in a relatively hot transition
region between a hot flow and a thin disk. 
The truncation radii of $\sim100 r_S$ radii inferred by 
Quataert et al. (1999) to fit
the optical/UV to X-ray SED is somewhat larger than the best-fit inner
radii inferred from the diskline model fits to the iron K$\alpha$ line
profile. However, is has been noted the  if a standard disk is connected to an
ADAF across a narrow transition layer, the gas near the transition
radius must rotate with a super-Keplerian angular velocity
(Abramowicz, Igumenshchev, \& Lasota 1998) 
This means $v_{FWHM}{\rm (Fe~K\alpha)} > c ({r_g}/{r_{tr}})$ or $r_{tr} >
r_{in}\sim ({c^2}/{v_{FWHM}^2}) r_g$.  Additionally, the transition region,
being hot ($\sim10^8 {\rm~K}$), may not contribute the optical/UV
emission which determined the truncation radius quoted by Quataert 
et al. (1999).
Note also that ADAFs are likely to develop strong outflows
(Blandford \& Begelman 1999) or convection (Quataert \& Gruzinov 1999;
Narayan et al. 2000) which are likely to change the dynamics 
of the transition region.

\subsection{Alternative scenario -- multiple narrow lines?}
The broad iron line components of NGC~4579 and M~81 can also be described in
terms of multiple narrow emission lines from ionized iron, although with a
lesser statistical significance. Such lines can 
arise from  ($i$) hot thermal plasma, if present, outside the accretion disk
e.g host galaxy, or 
 ($ii$) a photo-ionized accretion disk with flat emissivity law extending down
to the last stable orbit (Yaqoob et al. 2003). The former process has already
been ruled out  due to the variability of the iron line from NGC~4579  
(Terashima et al. 2000). The later possibility is unlikely due to a 
low ionization parameter (see $\S 5.3$). Thus the case for multiple narrow 
lines for the iron K$\alpha$ complexes of NGC~4579 and M~81 is poor at
present.

\section{Conclusions}
Using the \xmm{} EPIC observations, we detected complex iron line
profiles from two LLAGN NGC~4579 and M~81. The line profiles are well
described by a combination of a narrow neutral line and a broad
Gaussian (FWHM $\sim 20000\kms$) or an accretion-disk line arising
from highly ionized (He- or H-like) iron. The line profiles from both
the AGN are remarkably similar except that the neutral iron K$\alpha$
line from M~81 is much weaker than that that from NGC~4579.  The main
result of this work is that the broad and highly ionized iron
K$\alpha$ lines from two LLAGNs NGC~4579 and M~81 are consistent with
accretion disks whose inner edges are restricted to large radii
$r_{in} \sim 100r_g$.

\acknowledgements We thank an anonymous referee for the comments and 
suggestions on this paper. This work is based on
observations obtained with \xmm{}, an ESA science mission with
instruments and contributions directly funded by ESA Member States and
the USA (NASA). REG acknowledges NASA award NAG5-9902 in support of
his Mission Scientist position on \xmm{}.

\clearpage

\clearpage

\begin{table*}
\caption{Best-fit spectral model parameters for NGC~4579 and M~81.}
\label{tab1}
{\small
\begin{tabular}{lccccccc}
\tableline\tableline
Comp. & Parameter\tablenotemark{a}    & \multicolumn{2}{c}{NGC~4579} & \multicolumn{2}{c}{M~81}
\\
 &  & model 1\tablenotemark{b}  & model 2\tablenotemark{b} & model 1\tablenotemark{b} & model 2\tablenotemark{b} \\ \tableline
PL   & $\Gamma_X$ & $1.77_{-0.07}^{+0.06}$ & $1.77_{-0.07}^{+0.06}$ & $1.86_{-0.02}^{+0.02}$ &  $1.86_{-0.02}^{+0.02}$ \\
   & $n_{PL}$\tablenotemark{c} & $1.02_{-0.09}^{+0.10}$ &  $1.02_{-0.09}^{+0.11}$ & $3.6_{-0.1}^{+0.1}$ & $3.6_{-0.1}^{+0.1}$ \\
Gaussian & $E_{K\alpha,n}$($\kev$) & $6.40$(f)  &  $6.40$(f) & $6.40$(f) & $6.40$(f) \\
  & $\sigma$ ($\kev$) &  0.00(f) & 0.00(f) & 0.00(f) & 0.00(f) \\
  & $f\tablenotemark{d}$  & $4.7_{-2.5}^{+2.3}$ &  $4.3_{-1.9}^{+2.5}$ & $3.7_{-1.9}^{+1.6}$ & $4.2_{-2.3}^{+1.6}$    \\
  & $EW$ ($\ev$) & 106 &  96 & 31 & 35  \\
 Gaussian & $E_{K\alpha,b}$($\kev$) & $6.79_{-0.13}^{+0.11}$ &  -- & $6.79_{-0.07}^{+0.06}$ & --  \\
  & $\sigma$ ($\ev$) & $231_{-97}^{+102}$ & -- & $188_{-57}^{+76}$  & --\\
  & $f\tablenotemark{d}$  & $9.9_{-3.8}^{+4.4}$ &  -- & $10.4_{-2.6}^{+3.2}$  & -- \\
  & $EW$ ($\ev$) & 287 & -- & 101 & -- \\
DL & $E_{diskline}$($\kev$) & -- &  $6.78_{-0.09}^{+0.06}$ & -- &  $6.84_{-0.07}^{+0.04}$ \\
  & $q$ & -- & 2.5 & -- &  2.5(f) \\
  & $r_i/r_g$ & -- & $53_{-22}^{+137}$ & -- & $104_{-61}^{+110}$  \\
  & $r_o/r_g$ & --  & 1000(f) & -- & 1000(f) \\
  & $i$ & --  & 30(f) & -- & 30(f)  \\
  & $f\tablenotemark{d}$ & -- &  $10.0_{-3.6}^{+3.4}$ & -- & $10.1_{-2.3}^{+3.0}$
 \\
  & $EW_D$($\ev$) & --  & 302 & -- & 104 \\
Total  & $\chi^2_{min}$/dof &  242.5/249 & 240.6/249 & 299.3/300 & 296.3/300    \\ \tableline
\end{tabular}}
\tablenotetext{a}{(f) indicates a fixed parameter value.}
\tablenotetext{b}{model 1: absorbed power law $+$ narrow Gaussian $+$ broad Gaussian; model 2: absorbed power law $+$ narrow Gaussian $+$ diskline.}
\tablenotetext{c}{$10^{-3}{\rm~photons~cm^{-2}~s^{-1}~keV^{-1}}$ at $1\kev$.}
\tablenotetext{d}{$10^{-6}{\rm~photons~cm^{-2}~s^{-1}}$.}
\end{table*}

\end{document}